\documentclass[prl,twocolumn,superscriptaddress,showpacs]{revtex4}
\usepackage{graphicx}
\usepackage{amssymb}
\usepackage{amsmath}
\usepackage{color}
\usepackage{hyperref}

\newcommand{\equref}[1]{Eq.~(\ref{#1})}
\newcommand{\bitem}{\begin{itemize}}
\newcommand{\eitem}{\end{itemize}}
\newcommand{\benum}{\begin{enumerate}}
\newcommand{\eenum}{\end{enumerate}}
\newcommand{\btab}[1]{\begin{tabular}{#1}}
\newcommand{\etab}{\end{tabular}}
\newcommand{\btabn}[1]{\begin{tabular}{#1}}
\newcommand{\etabn}{\end{tabular}}
\newcommand{\beq}{\begin{equation}}
\newcommand{\eeq}{\end{equation}}
\newcommand{\beqn}{\begin{equation*}}
\newcommand{\eeqn}{\end{equation*}}
\newcommand{\bsplit}{\begin{split}}
\newcommand{\esplit}{\end{split}}
\newcommand{\Uwall}{\ensuremath{U_{\mathrm{wall}}}}
\newcommand{\gammadotav}{\ensuremath{\dot{\gamma}_{\mathrm{av}}}}

\newcommand{\gammadot}{\ensuremath{\dot{\gamma}}}

\begin{document}

\title{Heterogeneous shear in hard sphere glasses}
\author{Suvendu Mandal}
\affiliation{Interdisciplinary Centre for Advanced Materials Simulation (ICAMS), Ruhr-Universit\"at Bochum, Stiepeler Strasse 129, 44801 Bochum, Germany}
\affiliation{Max-Planck Institut f\"ur Eisenforschung, Max-Planck Str.~1, 40237 D\"usseldorf, Germany}
\author{Markus Gross}
\affiliation{Interdisciplinary Centre for Advanced Materials Simulation (ICAMS), Ruhr-Universit\"at Bochum, Stiepeler Strasse 129, 44801 Bochum, Germany}
\author{Dierk Raabe}
\affiliation{Max-Planck Institut f\"ur Eisenforschung, Max-Planck Str.~1, 40237 D\"usseldorf, Germany}
\author{Fathollah Varnik}
\email{fathollah.varnik@rub.de}
\affiliation{Interdisciplinary Centre for Advanced Materials Simulation (ICAMS), Ruhr-Universit\"at Bochum, Stiepeler Strasse 129, 44801 Bochum, Germany}
\affiliation{Max-Planck Institut f\"ur Eisenforschung, Max-Planck Strasse~1, 40237 D\"usseldorf, Germany}

\begin{abstract}
There is growing evidence that the flow of driven amorphous solids is not homogeneous, even if the macroscopic stress is constant across the system. Via event driven molecular dynamics simulations of a hard sphere glass, we provide the first direct evidence for a correlation between the fluctuations of the local volume-fraction and the fluctuations of the local shear rate. Higher shear rates do preferentially occur at regions of lower density and vice versa. The temporal behavior of fluctuations is governed by a characteristic time scale, which, when measured in units of strain, is independent of shear rate in the investigated range. Interestingly, the correlation volume is also roughly constant for the same range of shear rates. A possible connection between these two observations is discussed.
\end{abstract}

\maketitle

\paragraph{Introduction.---}
Heterogeneous flow and shear banding are central to the rheology of complex fluids and are widely observed in many industrial and natural materials, such as foams, emulsions, pastes, or even rocks \cite{olmsted_rheol_2008,schall_review2010}. Despite its ubiquitous appearance, many aspects of this phenomenon are still not well-understood. In the simplest case, shear banding can be captured by a nonmonotonic dependence of the shear stress, $\sigma$, on the shear rate, $\gammadot$ \cite{FieldingOlmstedCates,Fielding2009}. For certain complex fluids, like colloidal gels, shear banding can be associated to the competition between a structural phase transition 
and a shear
\cite{Moller}.
However, in other systems, such as dense hard sphere (HS) colloidal suspensions \cite{Besseling_SCC} and granular materials \cite{Losert2000}, flow heterogeneity is often observed without such accompanying structural changes. The rheological response of these systems is essentially determined by the competition between an inherent slow dynamics and the acceleration caused by the external drive \cite{Sollich1997,Berthier2002}. This may lead to a spatially and temporally heterogeneous flow if the system is close to the yielding threshold \cite{VarnikPRL_03,VarnikJCP_static_yields,Varnik2008}.

Recently, it has been proposed that flow localization in dense hard sphere suspensions can be alternatively rationalized in terms of shear-concentration coupling (SCC) \cite{Besseling_SCC}, a well-known feedback mechanism for flow instability in complex fluids \cite{SchmittPRE95_banding}. In this model, regions of high (low) shear rate are associated with high (low) diffusivity, giving rise to a shear-induced flux from high to low shear-rate regions. This leads to an increase (decrease) of density and thus viscosity in low (high) shear-rate regions, thereby enhancing shear-rate fluctuations further. This, in turn, enhances the migration of particles from high to low shear-rate regions, unless balanced by diffusive counter flux.
As pointed out recently \cite{furukawa}, enhancement of density fluctuations and the associated formation of voids can also be essential for the occurrence of fracture.

However, while the experimental data in \cite{Besseling_SCC} are interpreted consistently within the proposed macroscopic picture, no test of the basic underlying assumptions, such as the presence of a correlation between shear-rate and concentration fluctuations or the growth of emergent velocity fluctuations, has been provided so far for colloidal hard sphere glasses. This is not surprising, as the relevant density fluctuations that trigger the initial instability are quite small and hardly accessible to experiments. Furthermore, the available experimental time window for the observation of velocity fluctuations is limited to a few hundred percent strain, thus making a temporal analysis rather difficult.

Here, we study these and related issues via event-driven molecular dynamics (MD) simulations of a polydisperse hard sphere system. It is explicitly shown that fluctuations of the local volume fraction, $\delta \phi$, are correlated to the fluctuations of the local shear rate $\delta \gammadot$. More precisely, a decrease of local density is accompanied by an increase of the local shear rate and vice versa. However, while the relative amplitude of shear-rate fluctuations increases upon decreasing shear rate, the magnitude of the fluctuations of volume fraction remain constant. The temporal behavior of the fluctuations is characterized by a dominant time scale which is identical for both observables $\delta \phi$ and $\delta \gammadot$. When measured in units of strain, this time scale is of the order of a few hundred percent strain and fairly independent of the imposed shear rate in the studied range. Interestingly, the correlation volume, as determined from the maximum of the four-point susceptibility of density fluctuations, $\chi_4=N[\left< f^2_q(t) \right> - \left< f_q(t) \right>^2]$, is also roughly constant in the investigated range ($N$ is the particle number and $f_q(t)=N^{-1}\sum^N_{i=1}\exp[ \mathrm{i} \mathbf{q} \cdot (\mathbf{r}_i(t) - \mathbf{r}_i(0) )]$ is the incoherent scattering function at wave vector $\mathbf{q}$). This suggests a possible link between an inhomogeneous shear and a dynamically heterogeneous environment with a given spatial correlation.

\paragraph{Simulation setup.---} We perform event driven MD simulations of a polydisperse (11\%) hard sphere system in 3D \cite{rapaport_book}. The volume fraction is varied from below to above the glass transition point, which, for the present polydisperse system, is located at a volume fraction of $\phi_{g}\approx 58.5\%$ \cite{Pussey2009}. The quiescent properties of this system have been studied extensively in \cite{Williams}. The temperature is fixed at $T=1$ via velocity rescaling. In all the simulations reported here, $N=7000$ particles~\footnote{We have explicitly tested that, in the investigated parameter range, the obtained results are independent of the particle number.} are placed in a random configuration between two walls. The walls are made of the same kind of particles as the bulk, but have infinite mass.
A constant shear is applied by moving the walls with velocities $\pm \Uwall$ in the $\pm x$ direction (planar Couette flow). The shear gradient points in the $+z$ direction. The (overall) shear rate is defined as $\gammadotav=2\Uwall/L_Z$, where $L_Z$ is the distance between the walls. Local quantities, such as velocity and density profiles, are computed as an average over particles within distinct layers of finite width parallel to the walls. Note that, for simplicity of notation, we will drop the subscript ``av.'' Local quantities will be identified by their arguments, e.g., $\gammadot(z)$ and $\phi(z)$.

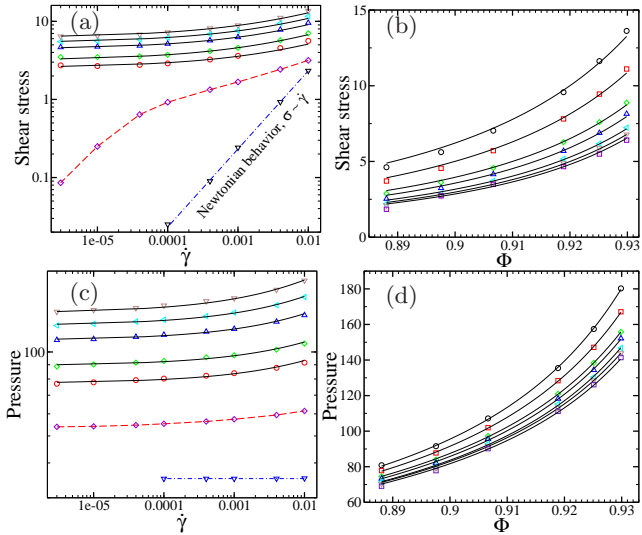
\begin{figure}[t]
\includegraphics*[height=3.5cm]{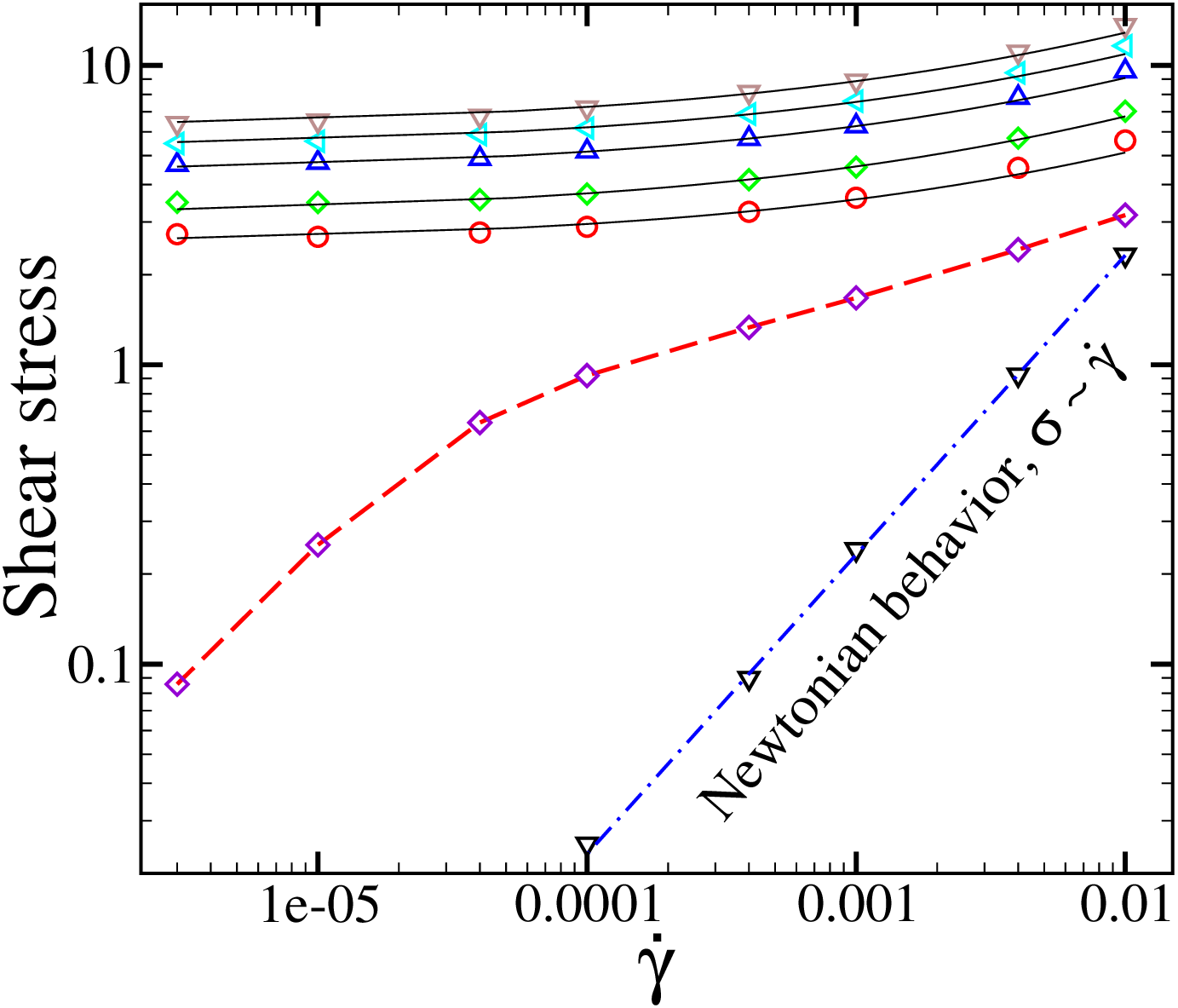}\hfil
\includegraphics*[height=3.5cm]{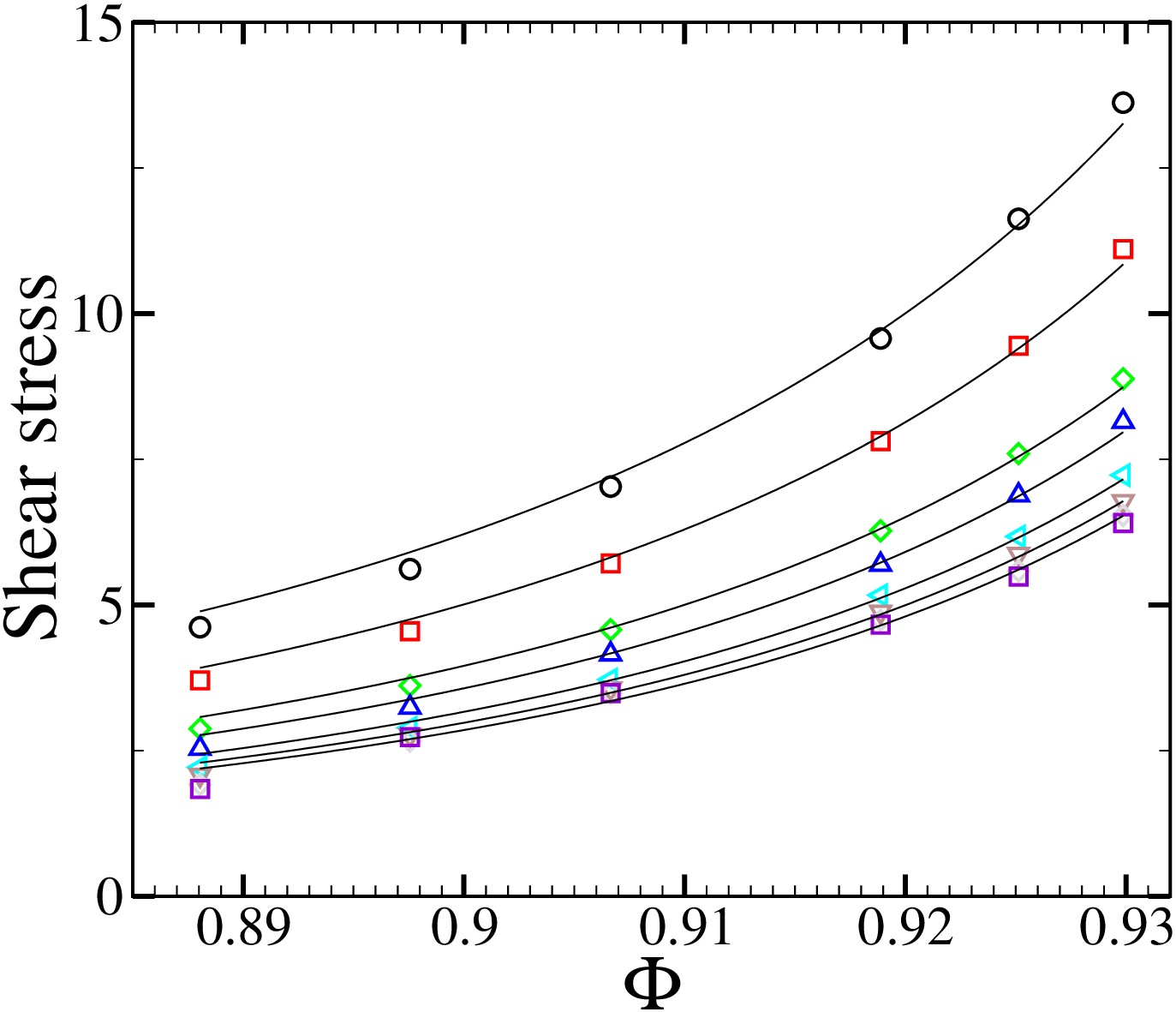}
\includegraphics*[height=3.5cm]{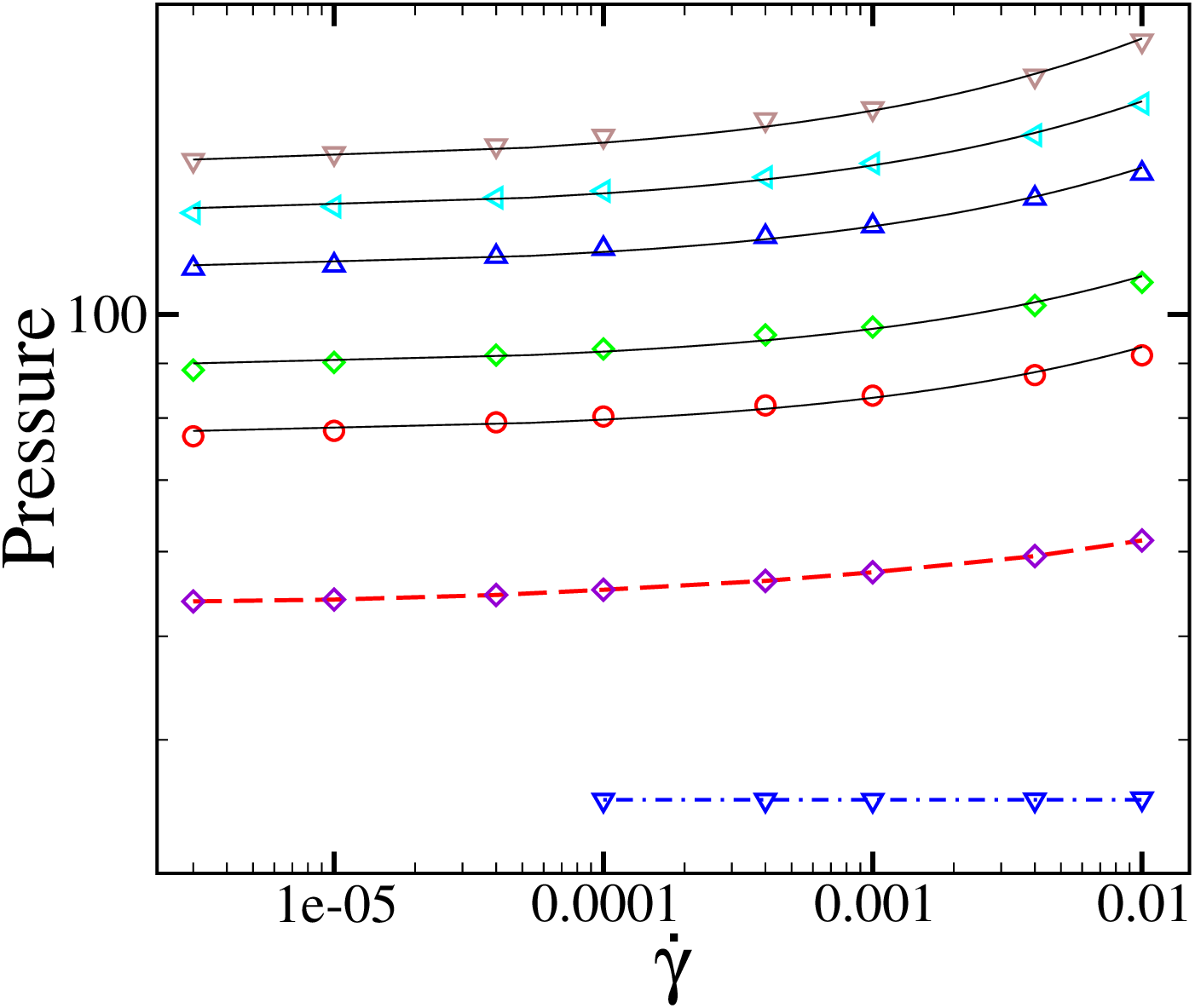}\hfil
\includegraphics*[height=3.5cm]{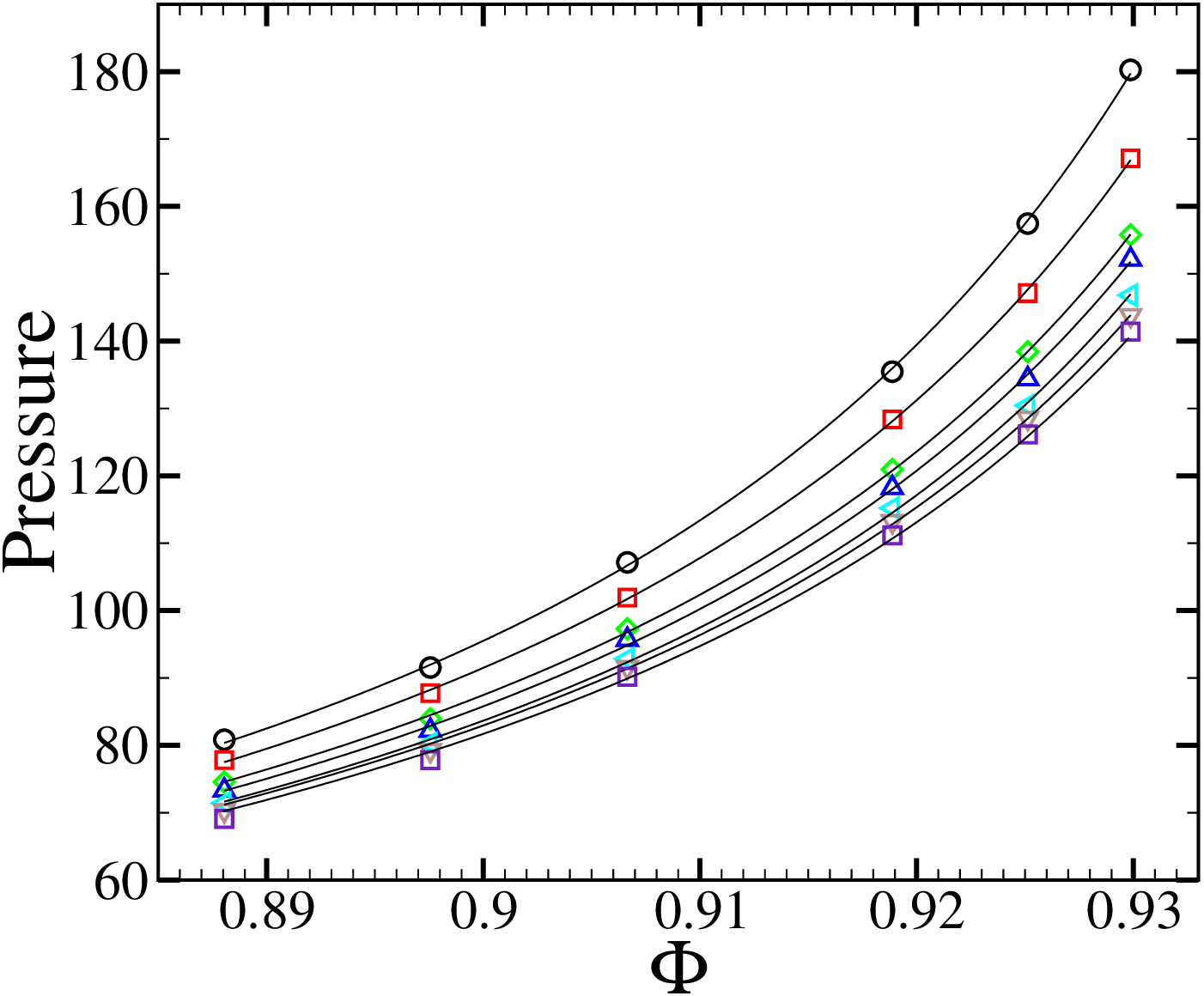}
\unitlength=1mm
\begin{picture}(0,0)
\put(-77,67){(a)}
\put(-35,66.5){(b)}
\put(-77,31){(c)}
\put(-35,30.5){(d)}
\end{picture}
\caption{Shear stress $\sigma$ and particle pressure $\Pi$ versus (a),(c) imposed shear rate $\gammadot$ and (b),(d) reduced volume fraction $\Phi$. Solid lines are fits to (a),(b) \equref{eq:Sigmaform} with parameters $p \simeq 2.355$, $n=0.4$, $\sigma_0=0.0119$, and $A=18-35$ and to (c),(d) \equref{eq:Piform} with parameters $r=3.8-4.1$, $m=0.4$, $\Pi_0 \simeq 8.4-10.05$, and $B=0.0015-0.007$. From bottom to top, different curves correspond to (a),(c) $\Phi=0.8060 (\phi=0.5400)$, $0.8657 (0.5800)$, $0.8970 (0.6014)$, $0.9067 (0.6075)$, $0.9189 (0.6157)$, $0.9251 (0.6198)$ and $0.9299 (0.6230)$ and (b),(d) $\gammadot=10^{-5}, 4 \times 10^{-5}, 10^{-4}, 4 \times 10^{-4}, 10^{-3}, 4 \times 10^{-3}$ and $10^{-2}$.}
\label{fig:rheology}
\end{figure}

\paragraph{Theoretical description.---}
Central to the theory of shear-concentration coupling are a non-Newtonian constitutive relation for the shear stress and the notion of a nonequilibrium particle pressure \cite{SchmittPRE95_banding}. Both quantities are coupled through the Navier-Stokes equations. In Figs.~\ref{fig:rheology}(a) and \ref{fig:rheology}(b), we show the dependence of the shear stress on shear rate and volume fraction in the yield stress regime, as obtained from our simulations. In the glassy phase ($\phi > 0.585 \leftrightarrow \Phi > 0.873$), our data can be well described by a Herschel-Bulkley expression
\begin{equation}
\sigma=\frac{\sigma_0}{(1-\Phi)^{p}}[1+s(\Phi){\rm \dot\gamma}^n], ~~s(\Phi)=A(1-\Phi)^n\,,
\label{eq:Sigmaform}
\end{equation}
where $p$, $n$, $\sigma_0$ and $A$ are constants to be determined by a fit. $\Phi=\phi/\phi_m$ is the reduced volume fraction ($\phi_m=0.67$ corresponds to random close packing). The first term in \equref{eq:Sigmaform} represents the dynamic yield stress \cite{Varnik2006}, while the second term accounts for the effect of shear. The dynamic yield stress can be naturally associated with shear heterogeneity, if the rigid regions are understood to be locally below the yielding threshold, while the liquidlike regions are above \cite{CoussotPRL02_shearbanding, VarnikPRL_03, schall_review2010, Besseling_SCC}. However, it does not explain the mechanism of the initial flow instability.

The particle pressure obtained from our simulations for different densities and shear rates is shown in Figs.~\ref{fig:rheology}(c) and \ref{fig:rheology}(d). Noting that the pressure data show similar behavior to the shear stress, it appears natural to write ($m$, $r$, $\Pi_0$ and $B$ are fit parameters)
\begin{equation}
\Pi=\frac{\Pi_0\Phi}{(1-\Phi)}[1+g(\Phi){\rm \dot\gamma}^m],~~g(\Phi)\equiv B(1-\Phi)^{1-r}.
\label{eq:Piform}
\end{equation}
The first term constitutes the pressure contribution responsible for ordinary concentration diffusion \cite{Brady1995}. The shear-rate dependence of the particle pressure is a manifestation of shear-induced particle migration, also known as ``dilatancy'' \cite{leighton_acrivos, Deboeuf, Haxton2011}.

Within SCC, the feedback responsible for heterogeneous flow occurs if an initial density excess leads, via 
$\partial_\phi \sigma$, to a locally lower shear rate, which in turn drives, via $\partial_{\gammadot}\Pi$, further particle migration towards the low shear region. The flow instability develops as soon as these fluxes cannot be overcome anymore by the concentration and viscous momentum diffusion, described by $\partial_\phi \Pi$ and $\partial_{\gammadot} \sigma$, respectively. A detailed calculation \cite{Besseling_SCC, SchmittPRE95_banding} shows that this happens if
\begin{equation}
F\equiv (\partial_{\dot\gamma} \Pi) (\partial_\Phi \sigma) / (\partial_\Phi \Pi) (\partial_{\dot\gamma} \sigma)>1.
\label{eq:modSchmittcrit}
\end{equation}
In the interesting lower shear-rate regime, $F$ reduces to
\begin{equation}
F \to {\rm \dot\gamma}^{m-n}\, \frac{mpg(\Phi)\Phi }{ns(\Phi)}\simeq
\frac{mpB\Phi}{nA(1-\Phi)^{r+n-1}} = F_0. \label{eq:F0}
\end{equation}
As seen from Figs.~\ref{fig:rheology}(a) and \ref{fig:rheology}(c), for the glassy phase in the limit of low shear rates, shear stress as well as pressure show quite a similar dependence on shear rate. Indeed, trying various fit procedures revealed that $m \approx n$ leads to consistent fit results for all the simulated data. This suggests that $F_0$ is practically independent on $\gammadot$. Using this information in $F(\dot\gamma_c,\Phi)=1$, the critical shear rate for the onset of instability is obtained:
\begin{eqnarray}
{\dot\gamma_c}(\Phi) \simeq \left[\frac{p}{r s(\Phi)[1-F(\gammadot \to \infty)]}\left(1-\frac{1}{F_0(\Phi)}
\right)\right]^{1/n}. 
\label{eq:gamma_c_low}
\end{eqnarray}
\color{black}

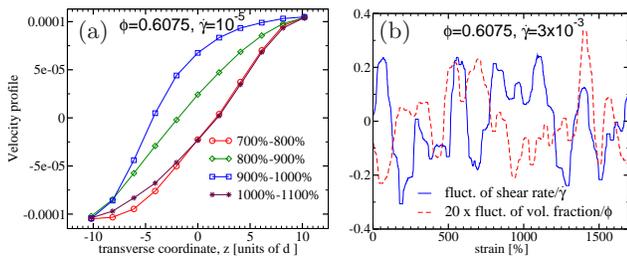
\begin{figure}[b]
\includegraphics*[height=3.3cm]{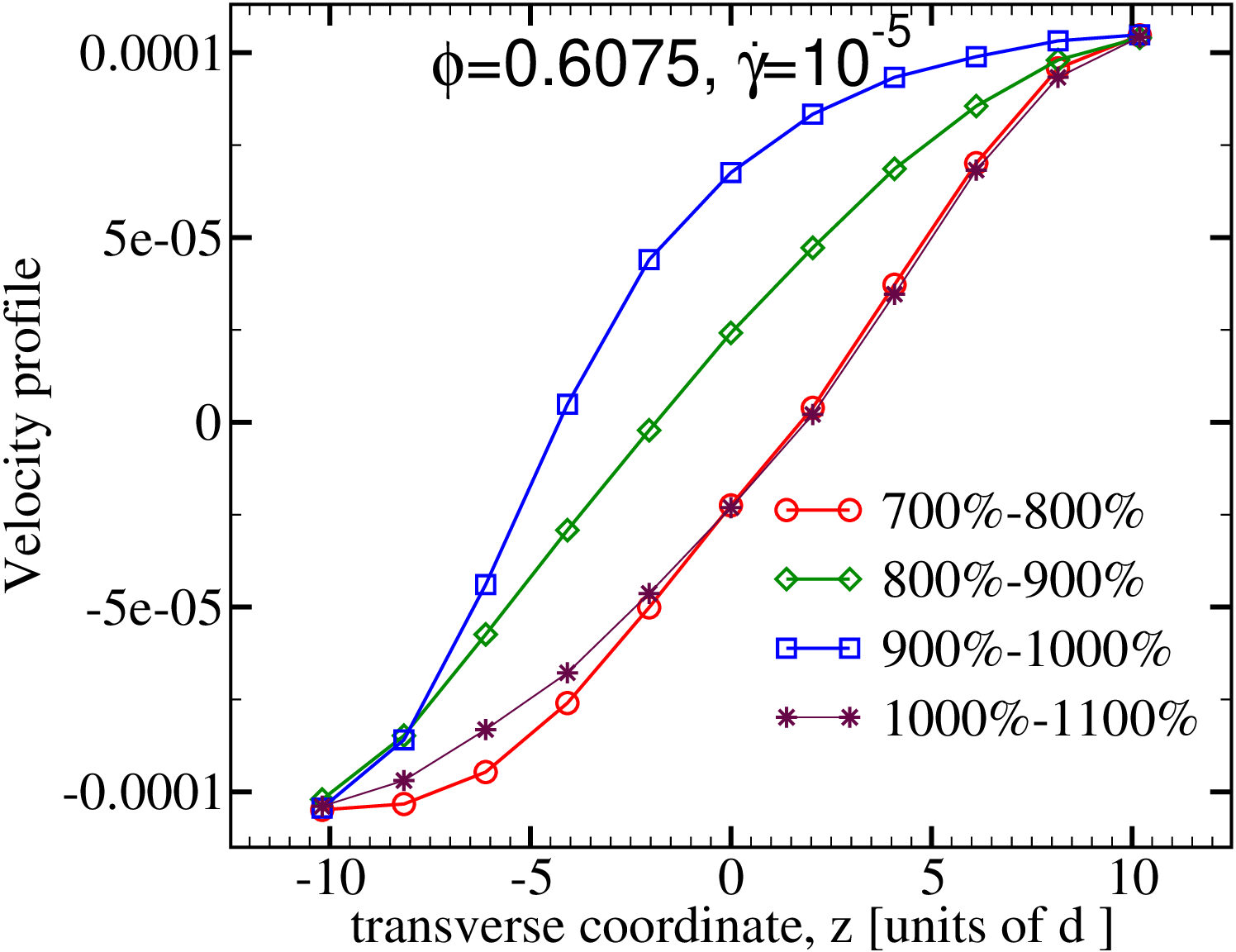}\hfill
\includegraphics[height=3.3cm,clip]{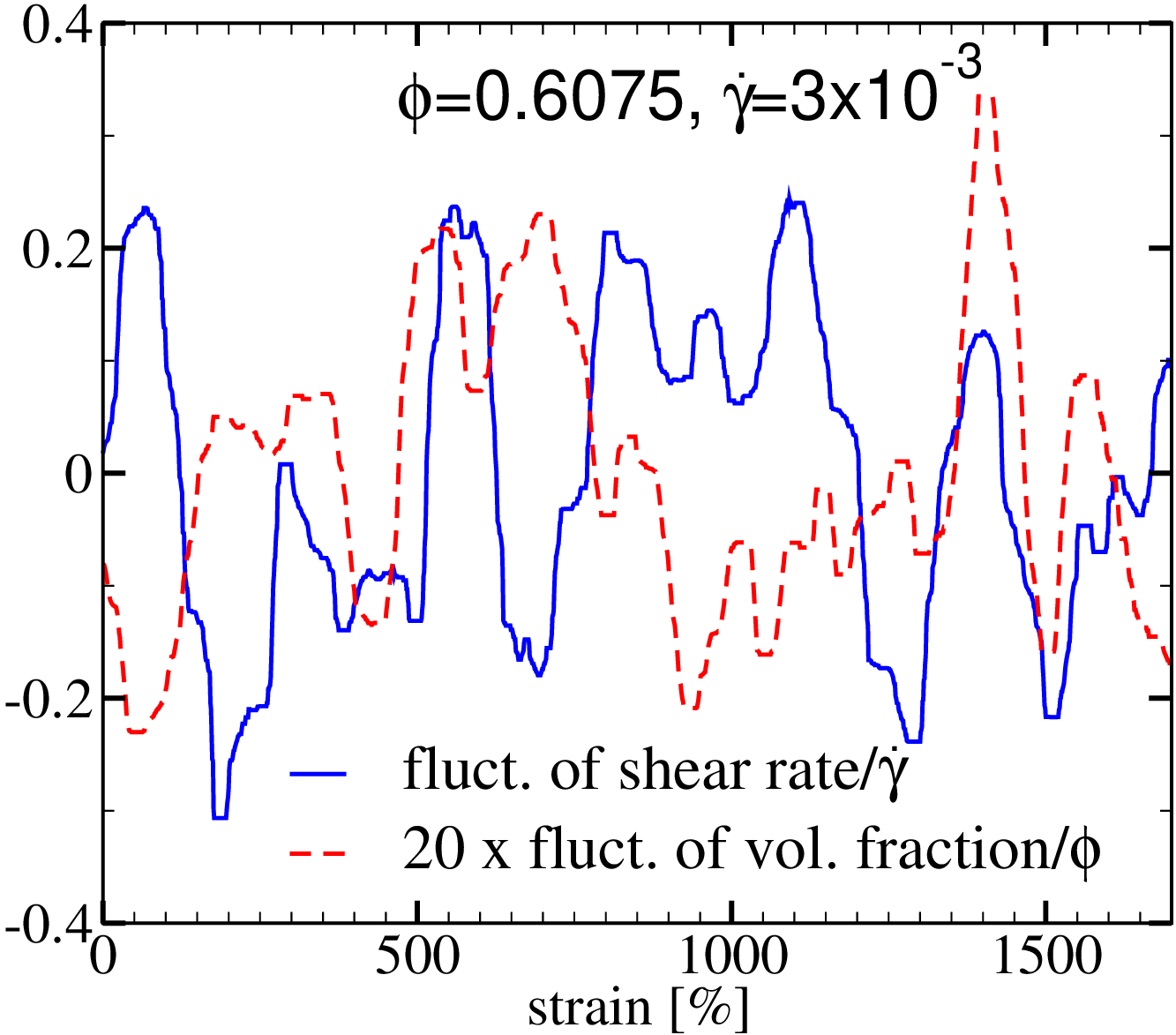}\hfill
\unitlength=1mm
\begin{picture}(0,0)
\put(-76,29){(a)}
\put(-36,29){(b)}
\end{picture}
\vspace{-0.3cm} 
\caption[]{
(a) Fluctuation of the velocity profile in a sheared HS glass. Each curve is computed as an average over the indicated strain interval (100\% strain corresponds to a time of $1/\gammadot$). (b) $\delta \gammadot/\gammadot$ and $\delta \phi/\phi$ versus time at the center of the simulation cell for a 300 times higher shear rate.}
\label{fig:hetflow}
\end{figure}

\paragraph{Heterogeneous shear.---}
Figure \ref{fig:hetflow}(a) shows the velocity profile at successive strain intervals, illustrating the occurrence of flow heterogeneity in the studied polydisperse HS model in the glassy phase ($\phi=0.6075$). Note the qualitative similarity to experiments of colloidal hard spheres \cite{Besseling_SCC} and to simulations of a binary Lenard-Jones glass \cite{VarnikPRL_03}. Here, we go a step further and perform a detailed survey of temporal fluctuations of both local shear rate, $\delta \gammadot = \gammadot(z) - \gammadot$, and volume fraction, $\delta \phi = \phi(z) - \phi$ [Fig.~\ref{fig:hetflow}(b)]. The data suggest that a positive $\delta \phi$ is often accompanied by a negative $\delta \gammadot$ and vice versa. Moreover, even though the temporal analysis is performed at considerably higher $\gammadot$ (due to limited computer time), the time scale of fluctuations in Figs.~\ref{fig:hetflow}(a) and \ref{fig:hetflow}(b) seems to be roughly the same in units of strain.

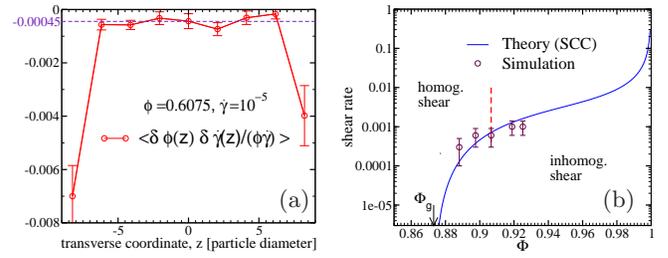
\begin{figure}[t]
\includegraphics*[height=3.3cm]{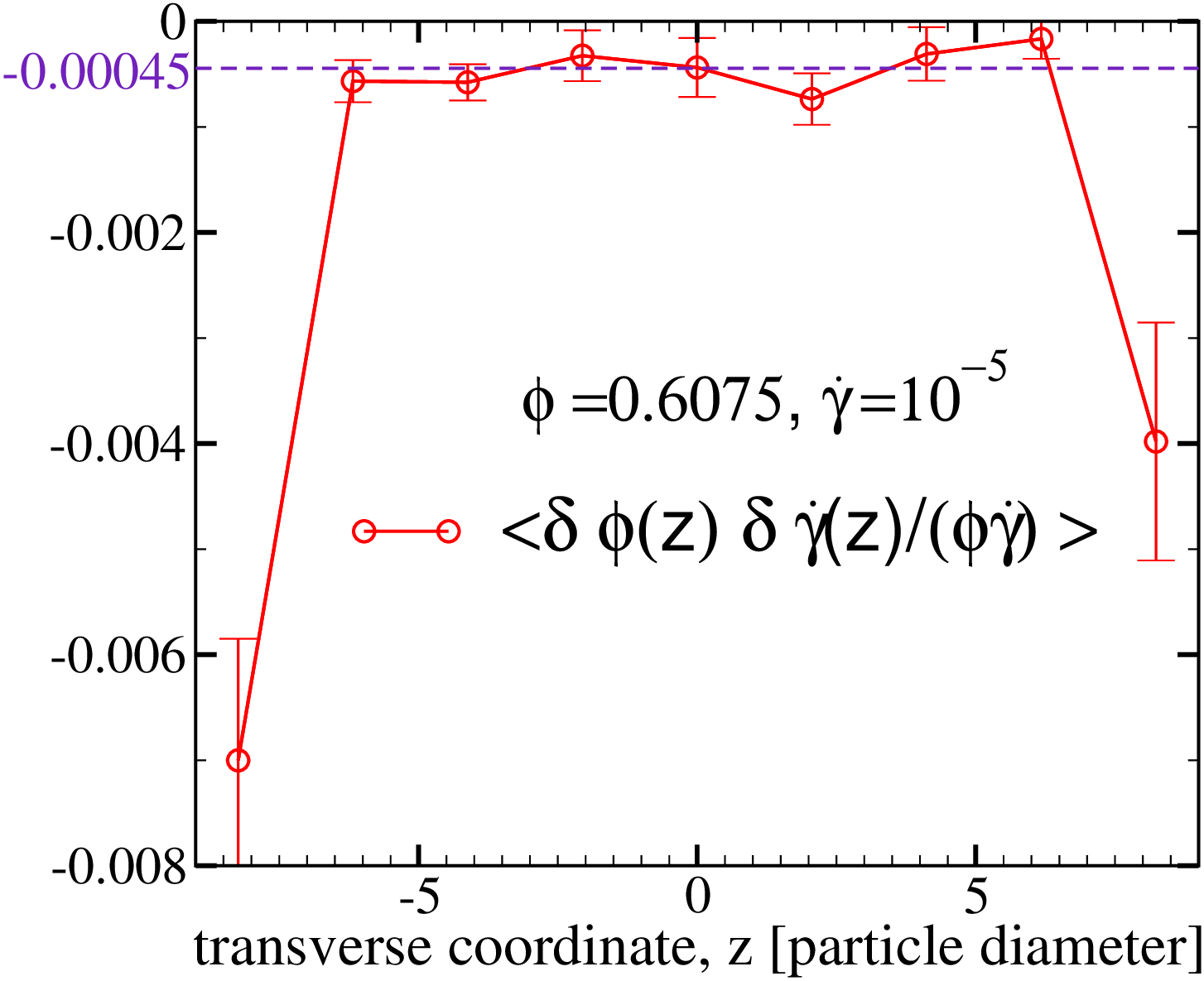}\hfill
\includegraphics*[height=3.3cm]{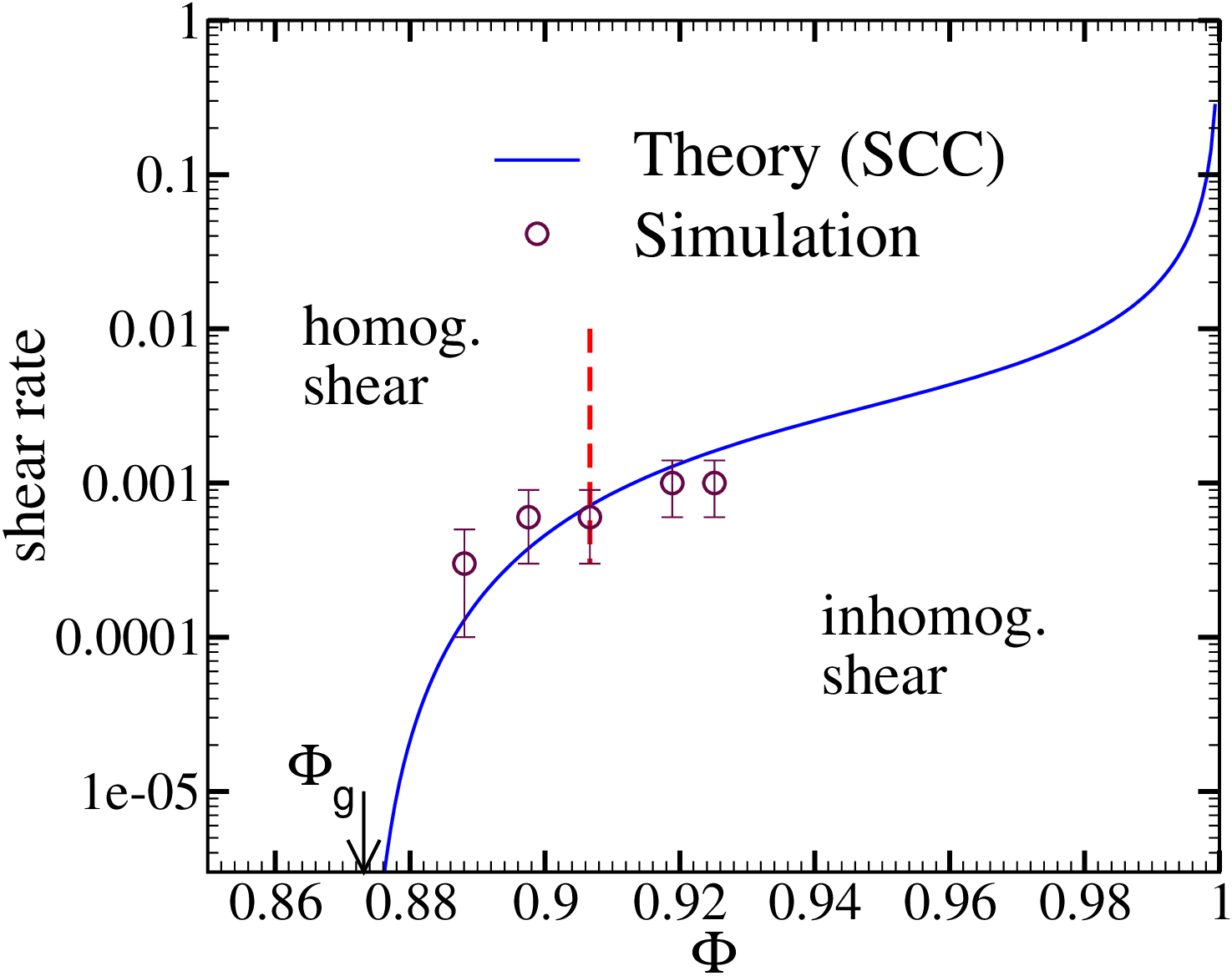}
\unitlength=1mm
\begin{picture}(0,0)
\put(-51,6){(a)}
\put(-8,6){(b)}
\end{picture}
\caption{(a) Correlation between fluctuations of local shear rate and volume fraction, determined within parallel layers of two-particle diameter thickness. (b) Comparing simulation results to \equref{eq:gamma_c_low} with $p=2.355$, $r=4.1$, $m,n=0.4$, $A=34.5$ and $B=0.007$. The dashed line marks the parameter range for which a detailed temporal analysis of fluctuations is performed (see Fig.~\ref{fig:omega}).}
\label{fig:instability}
\end{figure}

We compute the statistical average of the instantaneous correlations between fluctuations of the local shear rate and volume fraction, $C_{\phi\gammadot}(z)=\left< \delta \gammadot(z)  \delta \phi(z)/(\phi \gammadot) \right>$. A typical result on $C_{\gammadot\phi}(z)$ is shown in Fig.~\ref{fig:instability}(a) for an average shear rate of $\gammadot=10^{-5}$. We first note that most data points are distributed around a constant value indicated by a dashed line in the plot. This value is small but definitely nonzero (see the error bars, representing a standard deviation determined from independent runs). It is worth mentioning that the computational effort to obtain the present statistical accuracy is quite significant (40 independent runs, each with a duration of 60 days on a 3GHz CPU). The large values of $C_{\gammadot\phi}(z)$ at $z\approx \pm 9$ are a consequence of wall effects. Indeed, the volume fraction close to the walls is slightly (by about $1\%$) but systematically larger than the average volume fraction, leading to a corresponding decrease of wall shear rate. A detailed study of this issue will be presented elsewhere. For the reminder of this Letter, we will be concerned with bulk properties only.

\begin{figure}[b]
\includegraphics*[height=3.6cm]{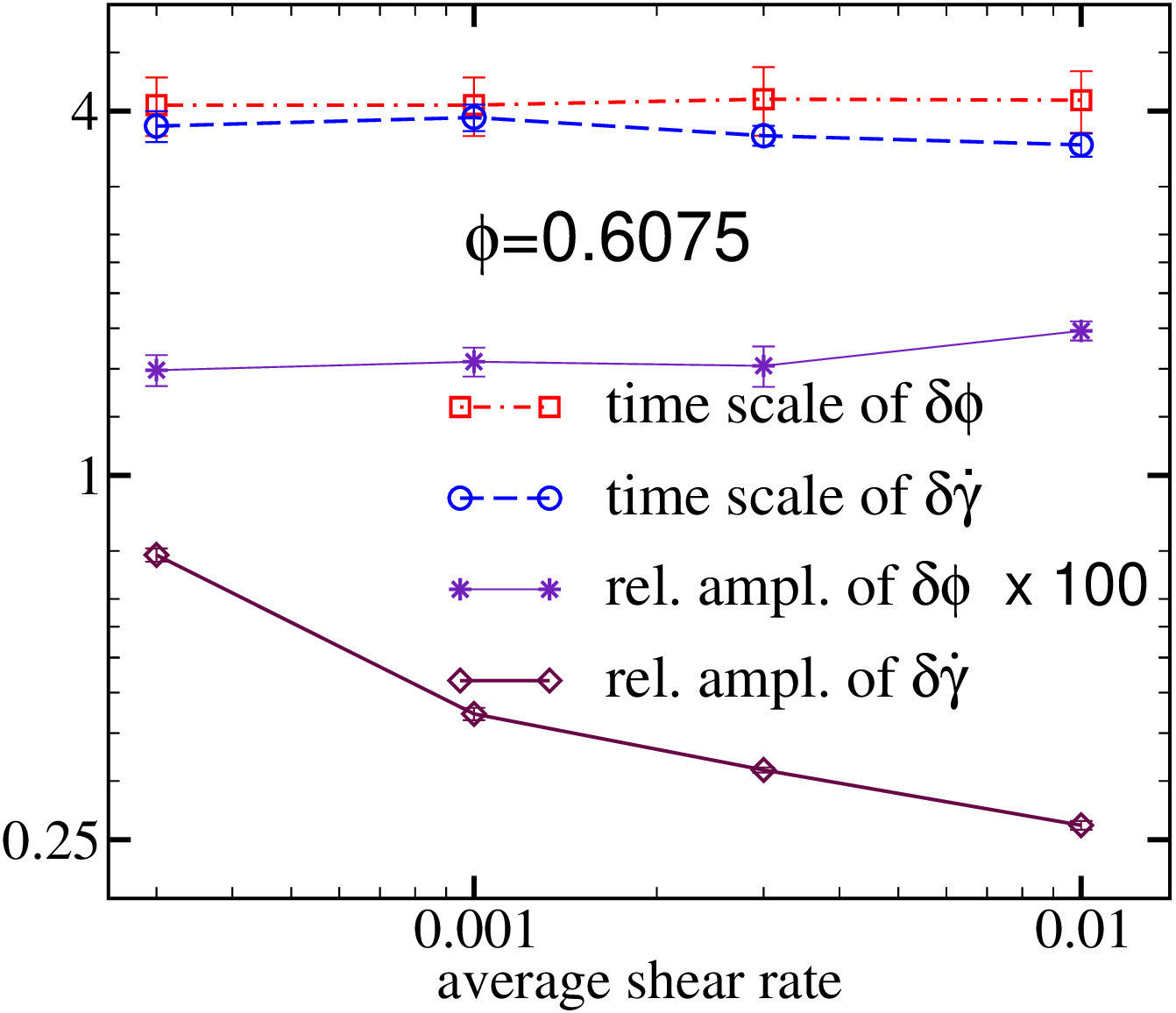}
\includegraphics*[height=3.6cm]{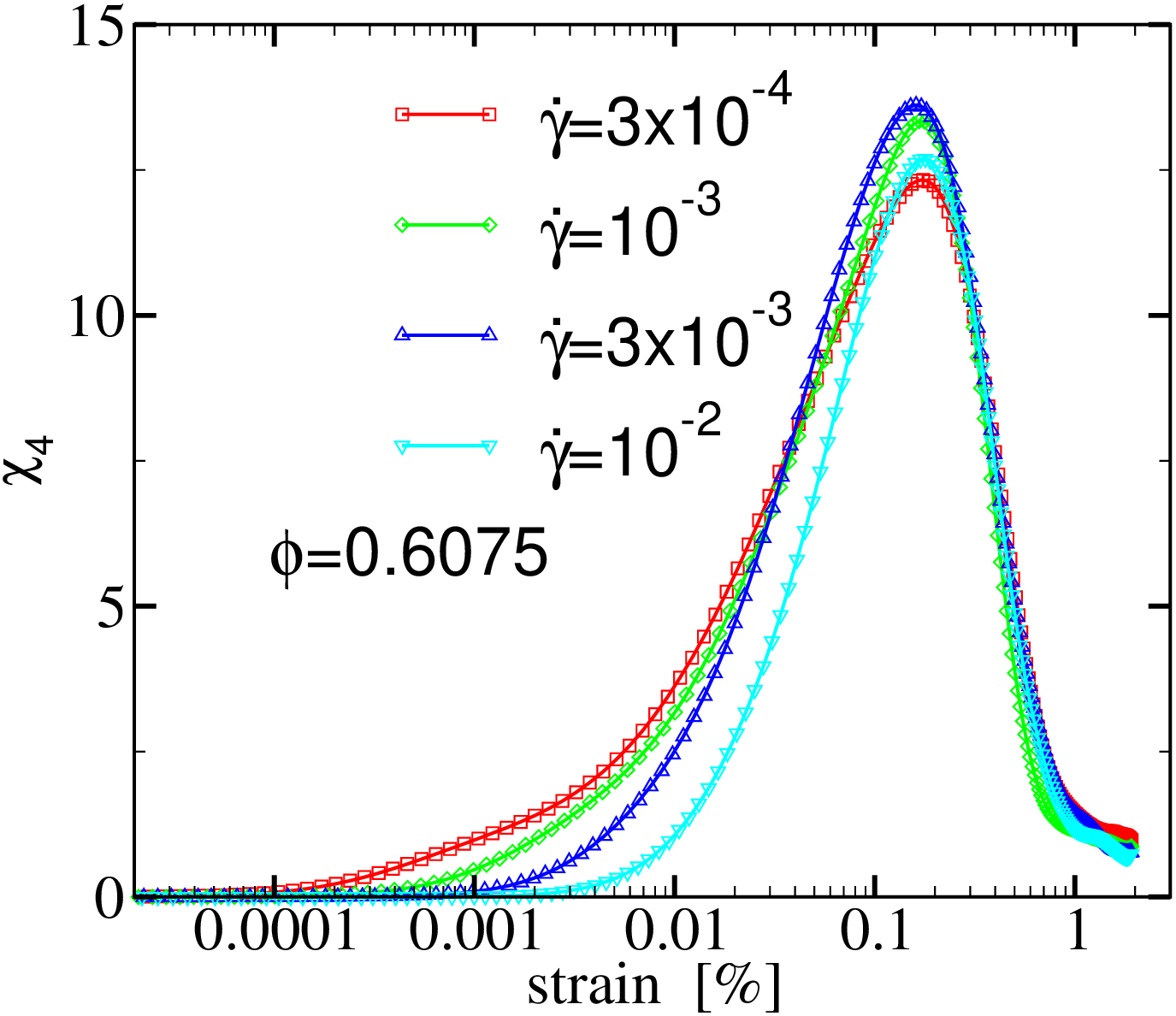}
\unitlength=1mm
\begin{picture}(0,0)
\put(-80,27){(a)}
\put(-36,31){(b)}
\end{picture}
\caption{(a) Dependence on shear rate of the relative amplitude and the characteristic time scale of fluctuations. (b) The function $\chi_4$ whose maximum is a measure of the correlation volume.}
\label{fig:omega}
\end{figure}
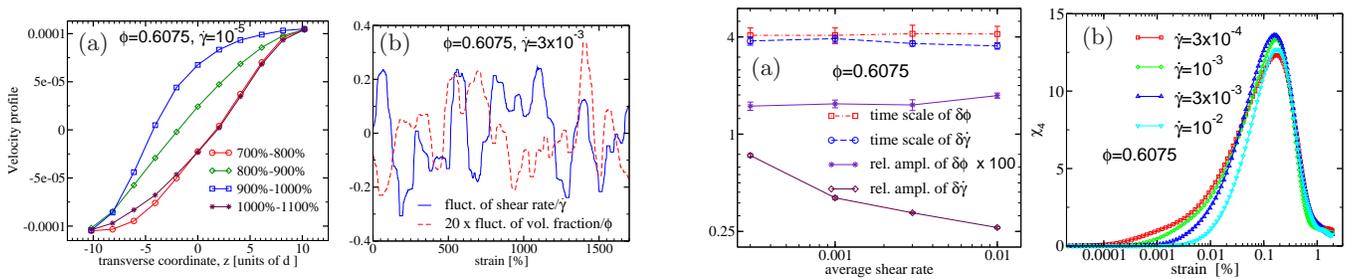

To test the stability phase diagram, Eq.~\equref{eq:gamma_c_low}, via our simulations, we identify a heterogeneous flow by requiring max$|(v(z)-\gammadot z) / \Uwall | >0.18$. As can be seen from Fig.~\ref{fig:instability}(b), simulation and theoretical predictions are in reasonable agreement. Thus, SCC seems to describe at least the onset of instability quite well for the present system. 

\paragraph{Time scale of fluctuations.---}
To elucidate the nature of the instability, we study layer-resolved fluctuations, $\delta\gammadot(z,t)$ and $\delta\phi(z,t)$, and determine, via a Fourier analysis, the peak frequency and the associated characteristic time scale of fluctuations at $\phi=0.6075$ for shear rates ranging from homogeneous to inhomogeneous flow regimes [dashed line in Fig.~\ref{fig:instability}(b)]. We first note that this time scale is independent of $z$ if the distance to the walls is above a few particles in diameter (data not shown). More importantly, it is practically identical for both the observables $\delta\gammadot$ and $\delta\phi$ [Fig.~\ref{fig:omega}(a)]. Moreover, when measured in the units of strain, it is independent of applied shear in the investigated range.
Using the same set of data, we also determine the relative amplitude of fluctuations as a function of shear rate [Fig.~\ref{fig:omega}(a)]. In marked contrast to the behavior of the characteristic time scale, this quantity depends on the specific observable. In particular, $\delta\gammadot/\gammadot$ increases significantly with decreasing imposed shear rate, while $\delta\phi/\phi$ is perfectly constant in the interesting limit of low shear rates, i.e., when entering the inhomogeneous flow regime.

The above observations have a number of consequences for a possible interpretation of flow heterogeneity. First, the presence of a correlation between fluctuations of local shear rate and volume fraction [Fig.\ref{fig:instability}(a)] and the analysis of the stability diagram [Fig.~\ref{fig:instability}(b)] are, at least in the first sight, in favor of SCC theory. However, the basic assumption of SCC theory is that fluctuations of shear rate give rise to fluctuations of volume fraction and vice versa. This implies that, as the instability is approached, \emph{both} $\delta \gammadot$ and $\delta\phi$ should grow to some extent. This is at odds with the data shown in Fig.~\ref{fig:omega}(a). To see this, we first note that, in contrast to the supercooled state, where a slight change of volume fraction can lead to considerable variations of viscosity as the glass transition is approached, the viscosity of a glass changes only little with volume fraction. This can easily be inferred from a comparison of shear stress for volume fractions below and above $\phi_{g}=0.585\, (\Phi_{g}=0.873)$ in Fig.~\ref{fig:rheology}(a). It is thus not plausible that tiny (in the sense of hardly detectable) changes of volume fraction can account for the observed increase of shear-rate fluctuation amplitude.

In this light, it is tempting to search for a link between the observed temporal behavior of $\delta\gammadot$ and $\delta\phi$ and dynamic heterogeneity. Without any claim for rigor, we start from the intuitive idea that some regions in the material are more mobile (i.e., appear more fluidlike) than others and therefore can support larger shear. These regions are not static. Rather, they continuously form and dissociate. The time scale of the fluctuations in the flow velocity is thus expected to reflect a time scale inherent to this microscopic dynamics. This inherent time scale should in turn be related to the characteristic size of these dynamically correlated regions. 
If this idea is consistent, the constant time scale~\footnote{Here we assume that, in the present case, the strain is the relevant unit of time (note that thermal fluctuations are negligible).} observed in our simulations would then imply a constant correlation volume. This is indeed borne out in Fig.~\ref{fig:omega}(b), where we plot $\chi_4$, whose maximum is a measure of the correlation volume \cite{Martens2011}, as a function of shear rate. We remark that dynamic correlations are not visible in the fluctuations of one-particle density so that, based on this idea, no conclusion can be made as to how local density fluctuations should behave. Therefore, the different behaviors of the fluctuation amplitudes of shear rate and density shown in Fig.~\ref{fig:omega}a are not a priori in conflict with this picture.

In conclusion, we find that, although the theory of shear-concentration coupling describes some aspects of flow heterogeneity in hard sphere glasses, serious inconsistencies remain, such as a significant increase of the relative amplitude of shear-rate fluctuations upon approaching the instability, as opposed to constant fluctuations of volume fraction. A more stringent test of the SCC theory would be to numerically solve the underlying Navier-Stokes equations using the experimental (or simulated) $\sigma(\gammadot, \phi)$ and  $\Pi(\gammadot, \phi)$ as input \cite{benzi_glasses}. An important question to be answered would be whether the original SCC theory predicts stable shear bands or a fluctuating behavior as observed in our simulations. Furthermore, keeping in mind that solvent effects on the system response are negligible in the limit of high $\phi$ and low $\gammadot$, it would also be interesting to work out the consequences of shear-concentration coupling by solving the \emph{compressible} Navier-Stokes equation (complemented by a non-Newtonian constitutive law) only for the colloids. When seeking for alternative routes, on the other hand, it must be kept in mind that, in its present form, the proposed SCC theory for a HS glass is a hydrodynamic approach with a fully local constitutive law. Dynamic correlations, however, give rise to nonlocal effects. These could possibly be taken into account (e.g., by means of a correlation length of some measure of ``fluidity'' \cite{Goyon2008}) by introducing generalized hydrodynamic equations.



\begin{acknowledgments}
S.M. is financially supported by the Max-Planck Society. We thank R.\ Besseling and A.\ Donev for useful discussions. Boundary-driven simulations are performed using the code by D. C.\ Rapaport \cite{rapaport_book}. We are also very grateful to M.\ Bannerman for adding polydispersity to DynamO \cite{DynamO}, which allowed us the study of homogeneous flow.
\end{acknowledgments}


\end{document}